\documentclass[11pt]{amsart}
\usepackage{wrapfig}
\usepackage{graphicx}

\newcommand{\noi}{\noindent}

\newcommand{\fr}{\frac}
\newcommand{\hb}{\hbar}

\newcommand{\vp}{{\bf p}}
\newcommand{\vP}{{\bf P}}
\newcommand{\vk}{{\bf k}}
\newcommand{\vr}{{\bf r}}
\newcommand{\ak}{{b_\vk}}
\newcommand{\aka}{{b^+_\vk}}
\newcommand{\vm}{{\bf m}}
\newcommand{\am}{{b_\vm}}
\newcommand{\ama}{{b^+_\vm}}

\newcommand{\om}{\omega}

\newcommand{\ran}{\rangle}
\newcommand{\lan}{\langle}

\newcommand{\del}{\delta}

\newcommand{\beg}{\begin{equation}}
\newcommand{\bg}{\begin{equation}}
\newcommand{\en}{\end{equation}}

\newcommand{\ba}{\begin{eqnarray}}
\newcommand{\bac}{\begin{array}{c}}
\newcommand{\eac}{\end{array}}
\newcommand{\bal}{\begin{eqnarray}{lcl}}
\newcommand{\eal}{\end{eqnarray}}
\newcommand{\ea}{\end{eqnarray}}

\newcommand{\dsum}{\displaystyle\sum}
\newcommand{\al}{\alpha}
\newcommand{\nn}{\nonumber}
\newcommand{\bt}{\begin{tabular}}
\newcommand{\et}{\end{tabular}}

\begin{document}

\thispagestyle{empty}
\begin{center}
\null\vspace{-1cm}
\vspace{1cm}
United Nations Educational Scientific and Cultural Organization\\
and\\
International Atomic Energy Agency\\
\medskip
THE ABDUS SALAM INTERNATIONAL CENTRE FOR THEORETICAL PHYSICS\\
\vspace{2cm}
{\bf INFINITELY IMPROVABLE UPPER BOUND ESTIMATES\\
FOR ACOUSTICAL POLARON GROUND STATE ENERGY}\\
\vspace{1.8cm}
N.N. Bogolubov, Jr.\footnote{nikolai\underline{ }bogolubov@hotmail.com}, A.V. Soldatov\footnote{soldatov@mi.ras.ru}\\
{\it V.A.Steklov Mathematical Institute of the Russian Academy of
Sciences,\\ 8 Gubkina Str., 119991 Moscow, Russia.}
\end{center}
\vspace{1cm}
\centerline{\bf Abstract}
\baselineskip=18pt
\bigskip

It was shown that an infinite convergent sequence of improving
non-increasing upper bounds to the ground state energy of a
slow-moving acoustical polaron  can be obtained by means of
generalized variational method. The proposed approach is
especially well-suited for massive analytical and numerical
computations of experimentally measurable properties of realistic
polarons and can be elaborated even further, without major
alterations, to allow for treatment of various polaron-type
models.

\noi {\bf Key words:} acoustic polaron, ground state, upper bound,
variational method

\noi {\bf PACS:}  71.38.-k, 71.38.Fp

\vfill
\begin{center}
MIRAMARE -- TRIESTE\\
October 2013\\
\end{center}
\vfill

\newpage

\section{The acoustical polaron model}

 A local change in the electronic state in a crystal leads to
the excitation of crystal lattice vibrations, i.e. the excitation
of phonons. And vice versa, any local change in the state of the
lattice ions alters the local electronic state.  This situation is
commonly referred to as an \lq\lq electron--phonon
interaction\rq\rq. This interaction manifests itself even at the
absolute zero of temperature, and results in a number of specific
microscopic and macroscopic phenomena such as, for example,
lattice polarization. When a conduction electron with
 band mass $m$ moves through the crystal, this state of
polarization can move together with it. This combined quantum
state of the moving electron and the accompanying polarization may
be considered as a quasiparticle with its own particular
characteristics, such as effective mass, total momentum, energy,
and maybe other quantum numbers describing the internal state of
the quasiparticle in the presence of an external magnetic field or
in the case of a very strong lattice polarization that causes
self-localization of the electron in the polarization well with
the appearance of discrete energy levels. Such a quasiparticle  is
usually called a \lq\lq polaron state\rq\rq\, or simply a \lq\lq
polaron\rq\rq.

The concept of the polaron was introduced first by L.D. Landau
\cite{Landau}, followed by much more detailed work by S.I. Pekar
\cite{Pekar1} who investigated the most essential properties of
stationary polaron in the limiting case of very intense
electron-phonon interaction, in the so-called adiabatic
approximation. Subsequently, Landau and Pekar \cite{LandauPekar}
investigated the self-energy and the effective mass of the polaron
for the adiabatic regime. Many other famous researchers have
contributed to the development of polaron theory later
\cite{Fr1,Fr2,Fr3,Feynman1,Badiab,BB1,LLP}.

A quantized polaron model for the case of an electron interacting
with longitudinal optical phonons, widely known as the Fr\"ohlich
polaron model, was introduced by H. Fr\"ohlich \cite{Fr3}. Since
then, a broad variety of  polaron-like models  has been devised on
its basis to account for the effects of the interaction of
electrons with other various types of phonons in crystals.  The
model under consideration is represented by the standard quantized
 acoustical polaron Hamiltonian

\beg H= \fr{\hat \vp^2}{2m}+\dsum_\vk \hb\om_{\vk}\aka\ak
+\dsum_\vk \tilde V_k\left(\aka e^{-i\vk\hat\vr}+  \ak
e^{i\vk\hat\vr}\right), \label{original}\en

\noi where $\om_{\vk}=sk$ is the frequency of the acoustical
phonons with $s$ being the velocity of sound,

\[ \tilde V_k=\left(\fr{4\pi
\alpha}{\tilde V}\right)^{1/2}\fr{\hb^2}{m}k^{1/2},
\]

\noi where $\tilde V$ is the volume of the crystal, and

\[ \alpha=\fr{D^2 m^2}{8\pi \rho \hb^3 s}\]

\noi is the dimensionless electron-phonon interaction constant
where $D$ is the deformation potential and $\rho$ the mass density
of the crystal.  The operators $\hat\vp$ and $\hat\vr$ stand for
the electron momentum and position coordinate quantum operators,
satisfying the usual commutation relations

\[
\quad [\hat p_i, \hat r_j]=-i\hb\del_{ij},
\]

\noi and the operators $b^+_{\vk}$, $b_{\vk}$, satisfying the
usual commutation relations

\[ [b_{\vk}, b_{\vk'}^+]=\del_{\vk\vk'},\quad [b_{\vk}, b_{\vk'}]=0,
\]

\noi are Bose operators of creation and annihilation of acoustical
phonons of energy $\hbar\om_{\vk}$ and wave vector $\vk$.

In the following it  will be convenient to express the energies in
units of $2ms^2$, the lengths in units of $\hb/2ms$ and the phonon
wave vectors in units of $2ms/\hb$ so that all variables will be
dimensionless. In this units the model (\ref{original}) takes the
form

\beg H= \hat \vp^2+\dsum_\vk k \aka\ak +\dsum_\vk V_k\left(\aka
e^{-i\vk\hat\vr}+  \ak e^{i\vk\hat\vr}\right),
\label{originaldl}\en

\noi with

\[  V_k=2\left(\fr{4\pi
\alpha}{ V}\right)^{1/2}k^{1/2},
\]

\noi where $V$ is dimensionless volume. In the course of the
calculations the sum over the phonon vectors $\sum_{\vk}$ will be
replaced finally by the integral $V/(2\pi)^3 \int d\vk$. In this
paper the so-called continuum polaron model (i.e. "large polaron")
is considered. But a finite cutoff at $k_0$, the boundary of the
first Brillouin zone in the phonon wave vector space,  is
introduced to account for the discreteness of the crystal lattice.
As usual, $k_0 \sim 1/a$, i.e. the inverse of the lattice
constant.

\section{Acoustical polaron ground state energy}

It is known that the polaron total momentum

\[
\hat \vP=\hat\vp+\dsum_{\vk}\vk \aka\ak
\]

\noi is a constant of the motion and commutes with the Hamiltonian
(\ref{original}). Therefore, it is possible to transform the
Hamiltonian to the representation in which $\hat \vP$ becomes a
"c"-number by means of the unitary transformation

\[
H\to\tilde H,\quad \tilde H= S^{-1}HS, \quad
S=\exp(-i\dsum_{\vk}\vk\hat\vr\aka\ak),
\]

\noi so that

\[ \tilde H= (\hat\vp - \dsum_{\vk}k \aka\ak
)^2+\dsum_\vk \vk\aka\ak +\dsum_\vk V_k\left(\aka + \ak \right),
\]

\noi or

\beg \tilde H= (\vP - \dsum_{\vk}\vk \aka\ak )^2+\dsum_\vk
k\aka\ak +\dsum_\vk V_k\left(\aka + \ak \right),
\label{corigtransf}\en

\noi in the $\hat\vp$ - representation where $\hat\vP$ becomes a
quantum "c"-number $\vP$, the value of the polaron total momentum,
and the Hamiltonian (\ref{corigtransf}) no longer contains the
electron coordinates. Another unitary transformation

\[
\tilde H\to {\mathcal H}(f),\quad {\mathcal H}(f)= U^{-1}\tilde
HU, \quad U=\exp\{\dsum_{\vk} f_{\vk}(\aka  - \ak )  \},
\]

\noi provides us with the Hamiltonian

\ba {\mathcal H}(f)= \left(\vP - \dsum_{\vk}\vk
(\aka+f_{\vk})(\ak+f_{\vk}) \right)^2+\nn\\ +\dsum_\vk k\aka\ak+
\dsum_{\vk}[kf_{\vk}+V_k](\aka+\ak)+
 2\dsum_{\vk} V_k f_{\vk}+\dsum_{\vk}k^2f^2_{\vk},
 \label{finalHam}\ea

\noi or, in a much more convenient albeit equivalent form,

\[ {\mathcal H}(f)={\mathcal H}_0(f)+{\mathcal H}_1(f), \]

\noi where

\[ {\mathcal H}_0(f)= P^2+\dsum_{\vk} k\aka\ak+(\dsum_{\vk} \vk
\aka\ak)^2 -\alpha', \]

\ba {\mathcal
H}_1(f)=\dsum_{\vk}[(k+k^2)f_{\vk}+V_k](\aka+\ak)+2\dsum_{\vk\vm}(\vk\cdot
{\bf m} ) f_{\vk}f_{\vm} \aka\am + \nn \\
+\dsum_{\vk\vm}(\vk\cdot {\bf m} ) f_{\vk}f_{\vm} (\aka\ama
+\ak\am) +
2\dsum_{\vk\vm}(\vk\cdot {\bf m} ) f_{\vk}(\ama\am\ak+\aka\ama\am  )-\nn\\
-2\dsum_{\vk}(\vP\cdot {\bf k} )(\aka+f_{\vk})(\ak+f_{\vk})+\nn\\
+2\dsum_{\vk\vm} (\vk\cdot\vm) f_{\vm}^2\aka\ak + 2\dsum_{\vk\vm}
(\vk\cdot\vm) f_{\vm}^2(\aka+\ak)+\dsum_{\vk\vm} (\vk\cdot\vm)
f_{\vm}^2f_{\vk}^2,
 \label{altfinalHam}\ea

 \noi
 and

 \[
-\alpha'=
2\dsum_{\vk}V_kf_{\vk}+\dsum_{\vk}(k+k^2)f^2_{\vk}+(\dsum_{\vk}f^2_{\vk}\vk)^2,
 \]

\noi which is just the sole Hamiltonian to be treated further on.

The ultimate goal is to find the lowest eigenvalue
$E_g(\alpha,\vP, k_0)$ of this Hamiltonian  corresponding to the
ground state energy of the slow-moving polaron for a given total
polaron momentum $\vP$. Then, the function $E_g(\alpha,\vP, k_0)$
could be expanded in powers of $\vP$ as

\[ E_g(\alpha, \vP, k_0)=E_g(\alpha,0,k_0) + \fr{P^2}{2m_{eff}} +
O(P\,^4), \]

\noi where $E_g(\alpha,0,k_0)$ is the ground state energy of the
polaron at rest and the coefficient $m_{eff}$ can be interpreted
as the polaron effective mass. In general spatially anisotropic
case,  the so-called inverse mass tensor

\[ \left( \fr{1}{m_{eff}}\right)_{ij}=\fr{\partial^2 E(\alpha,\vP,
k_0)}{\partial P_i\partial P_j}\Bigg|_{\vP=0} \]

\noi must be introduced instead of the scalar effective mass
parameter $m_{eff}$.

Extensive work has already been done to evaluate $E(\alpha, \vP,
k_0)$ directly through conventional perturbational calculations or
to find upper bound estimates for its value by means of
multitudinous variational methods. These approaches are beyond the
scope of this work. It is only worth mentioning that, as a rule,
perturbational schemes does not provide one with reliable error
bound estimates whilst the quality of upper bounds derived by
variational methods depends mostly on the choice of proper trial
states in any particular case and, being this way, these bounds
cannot be improved significantly, not to say infinitely, step by
step, through any regular scheme of calculations.

The purpose of the present research is to show that infinitely
improvable upper bounds for the ground state energy $E(\alpha,
\vP, k_0)$ can be obtained by generalized variational method
formulated for the first time in \cite{SolVar1} and later in
\cite{Kir1} in a different context.


\section{Generalized variational method}

It was proved in \cite{SolVar1} following ideas outlined in
\cite{PeetDevr}, and also found later in \cite{Kir1} by a
different approach, that for a quantum system represented by some
Hamiltonian $\hat H$ and any normalized trial state
$|\psi\rangle$,\, such that $\langle\psi|\psi\rangle =1$,

\[ E_g \le \min(a_1^{(n)},...,a_n^{(n)})\le \langle\psi| \hat
H|\psi\rangle, \]

\noi where the ordered by increase real numbers
$(a_1^{(n)},...,a_n^{(n)})$ are the roots of the n-th order
polynomial equation

\[ P_n(x)=\dsum_{i=0}^n X_ix^{n-i}=0, \]

\noi whereby $X_0\equiv 1$ and all the other coefficients $X_i$,\,
$1\le i\le n$ are provided by the system of $n$ linear equations

\[ {\mathcal M} {\bf X} +{\bf Y} =0, \]

\noi with

\[ Y_i=M_{2n-i}, \quad {\mathcal M}_{ij}=M_{2n-(i+j)}, \quad i,j
=1,2,...n, \]

\noi and

\[ M_m = \langle\psi| \hat H^m|\psi\rangle.\]

\noi It is assumed that all moments $M_m$ are finite. Moreover, it
was proved that a limit exists

\[ {\mathcal E}_g =\lim_{n\to\infty} \min(a_1^{(n)},...,a_n^{(n)}), \]

\noi and the following inequality holds

\[ \min(a_1^{(n+1)},...,a_{n+1}^{(n+1)}) \le
\min(a_1^{(n)},...,a_n^{(n)}). \]

\noi For example, at the first order

\[ E_g\le a_1^{(1)},\qquad  a_1^{(1)}= \langle\psi| \hat
H|\psi\rangle, \]

\noi and at the second order

\bg E_g\le \min(a_1^{(2)}, a_2^{(2)}) = \langle\psi| \hat
H|\psi\rangle+\fr{K_3}{2K_2}-\left[
\left(\fr{K_3}{2K_2}\right)^2+K_2 \right]^{1/2},\label{secorder}
\en

\[ a_1^{(2)} = \langle\psi| \hat
H|\psi\rangle+\fr{K_3}{2K_2}-\left[
\left(\fr{K_3}{2K_2}\right)^2+K_2 \right]^{1/2}, \]

\[  a_2^{(2)} = \langle\psi| \hat
H|\psi\rangle+\fr{K_3}{2K_2}+\left[
\left(\fr{K_3}{2K_2}\right)^2+K_2 \right]^{1/2}, \]

\noi where  $K_2$ and $K_3$ are the central moments

\[ K_2= \langle\psi| (\hat H -  \langle\psi| \hat H|\psi\rangle
)^2|\psi\rangle, \qquad K_3= \langle\psi| (\hat H - \langle\psi|
\hat H|\psi\rangle )^3|\psi\rangle. \]

\noi It is obvious  that the second order upper bound
(\ref{secorder}) would lie  below the first order upper bound for
most physically relevant quantum models and most reasonable
choices of the trial state $|\psi\rangle$.

\noi Moreover, if  $\langle\psi|E_g\rangle \ne0$ \,\,\,,
then\,\,\, $ \lim_{n\to\infty} \min(a_1^{(n)},...,a_n^{(n)})=E_g$.

Additionally, an excitation gap, should there happen to be any
discernable one in the spectrum, can be approximated at the $n$-th
order by the difference

\[ G_n= a_2^{(n)}- a_1^{(n)}.\]

\section{Infinitely improvable upper bounds for acoustical polaron at rest}

 For $\vP=0$, the function $f_{\vk}$ is  spherically symmetric, and the canonically  transformed  Fr\"ohlich polaron model
 (\ref{finalHam}) can be written down as

\ba {\mathcal H}(f)=  \dsum_{\vk} k\aka\ak+(\dsum_{\vk} \vk
\aka\ak)^2
-\alpha'+\nn \\
+ \dsum_{\vk}[(k+k^2)f_k+V_k](\aka+\ak)+2\dsum_{\vk\vm}(\vk\cdot
{\bf m} ) f_kf_m \aka\am + \nn \\
+\dsum_{\vk\vm}(\vk\cdot {\bf m} ) f_kf_m (\aka\ama +\ak\am) +
2\dsum_{\vk\vm}(\vk\cdot {\bf m} ) f_k(\ama\am\ak+\aka\ama\am
).\nn
 \ea

 \noi Let us choose phonon vacuum state $|0\rangle$ as a trial state $|\psi\ran$ for ${\mathcal
 H}(f)$,\, so that inequality

 \[ E_g(\alpha,0,k_0)\le \langle 0|{\mathcal H}(f)|0\rangle =2\dsum_{\vk}V_kf_k+\dsum_{\vk}(k+k^2)f^2_k
 \]

\noi holds, the right-hand side of which is minimized by

\[ f_k=-V_k/(k+k^2), \]

\noi and, eventually,

\bg  E_{g}(\alpha,0,k_0)\le E_{W}(\alpha,0,k_0)=
-\fr{4\alpha}{\pi}\left[2\ln[1+k_0]+k_0^2-2k_0\right].\label{LLP}
\en

\noi The  bound (\ref{LLP}) is precisely the upper bound derived
in \cite{SumiT,Peeters} by the Feynman path-integral method (note
that different units of the energies, lengths and wave vectors
were employed in \cite{Peeters}). Though this bound holds formally
for arbitrary strength of the electron-phonon interaction,  it is,
actually, the second order perturbation-theory result valid in the
weak-coupling limit. In order to derive better upper bounds at
higher orders of generalized variational method it is only
necessary to calculate moments $\lan 0|{\mathcal H}^m(f)|0\ran $
for sufficiently large integer exponents $m$. This can be easily
accomplished by means of the Wick theorem. The resulting
multitudinous products of integrals of the kind

\bg \int_0^{k_D}\fr{k^p dk}{(k+k^2)^q}, \qquad p,q -
\mbox{non-negative integers},\label{exint}\en

\noi can be evaluated  analytically wherever necessary as well as
all the concomitant integrals over the angular variables of the
corresponding wave vectors.

At the second order variational approximation (\ref{secorder})

\ba E_g(\alpha,0,k_0)\le E_{var}=
-\fr{4\alpha}{\pi}\left[2\ln[1+k_0]+k_0^2-2k_0\right]
+\fr{K_3}{2K_2}-\left[ \left(\fr{K_3}{2K_2}\right)^2+K_2
\right]^{1/2}, \label{SO}\ea

\noi where

\ba K_2= \fr{128\alpha^2}{3\pi^2} F_1^2(k_0), \label{K2}\ea

\ba K_3=\fr{256\alpha^2}{3\pi^2} F_1(k_0)F_2(k_0) +
\fr{4096\alpha^3}{9\pi^3} F_3^3(k_0), \label{K3}
 \ea

\ba F_1(k_0)= \left[-k_0+\fr{k_0^2}{2}+3\ln(1+k_0)+\fr{1}{1+k_0}-1
\right] \label{F1} \ea

\ba F_2(k_0)= \left[3k_0-k_0^2+
\fr{k_0^3}{3}-4\ln(1+k_0)-\fr{1}{1+k_0}+1 \right] \label{F2} \ea

\ba F_3(k_0)= \left[-4k_0+\fr{3k_0^2}{2}-
\fr{2k_0^3}{3}+\fr{k_0^4}{4}+5\ln(1+k_0)+\fr{1}{1+k_0}-1 \right]
\label{F3} \ea

 \noi It is instructive to compare this bound with the bound
obtained in \cite{Peeters} in the strong-coupling limit

\ba E_{SC}= - \fr{8\alpha}{3\pi}k_0^3 +
2\sqrt{2}\left[\fr{3\alpha}{5\pi}  \right]^{1/2}
k_0^{5/2}.\label{SC}\ea

\noi It is argued in \cite{Peeters}, that the strong coupling
region is defined by the condition  $\alpha>>15\pi/32 k_0$. Bounds
(\ref{LLP}),  (\ref{SO}) and (\ref{SC}) are plotted as functions
of $\alpha$ in Figs.1-4. for $k_0=0.5$, $k_0=1$, $k_0=2$ and
$k_0=3$. It is seen that for relatively small values of the
cut-off wave vector $k_0$ the bound $E_{var}$ is much lower than
the two other bounds for the whole region of the interaction
strength $\al$. As $k_0$ increases, the bound $E_{var}$ approaches
the weak-coupling limit bound $E_{W}$ from below for any fixed
value of $\al$. It seems that such asymptotic behavior of the
variational bound $E_{var}$ is typical for other polaron models
too, for example, for the "physical" Fr\"ohlich polaron model
\cite{SolCM}. Therefore, to obtain better bounds for larger values
of $k_0$, calculations at higher orders of the generalized
variational method are to be carried out.

\section{Infinitely improvable upper bounds for slow-moving acoustical polaron}

The same trial  state $|0\rangle$ can be employed in general case
$\vP \ne 0$ leading to inequality

\ba E_g(\alpha,\vP,k_0)\le \langle 0|{\mathcal H}(f)|0\rangle
=P^2+2\dsum_{\vk}V_kf_{\vk}+\dsum_{\vk}(k+k^2)f^2_{\vk}
-2\dsum_{\vk}(\vP\cdot {\bf k} )f^2_{\vk}+(\dsum_{\vk}
f^2_{\vk}\vk)^2,
 \label{gencase}\nn\ea

\noi the right-hand side of which is minimized by

\[ f_{\vk}=-V_k/[k-2\vk\cdot\vP(1-\eta)+k^2], \]

\noi where $\eta$ is defined self-consistently by the equation

\[ \eta\vP=
\dsum_{\vk}f^2_{\vk}\vk=\dsum_{\vk}V_k^2\vk/[k-2\vk\cdot\vP(1-\eta)+k^2]^2,\]

\noi or, alternatively, by

\[ \eta P^2
=\dsum_{\vk}V_k^2\vk\cdot\vP/[k-2\vk\cdot\vP(1-\eta)+k^2]^2.\]

\noi  The resulting upper bound

\ba E_g(\alpha,\vP,k_0)\le P^2(1-\eta)^2- \dsum_{\vk}V_k^2
\fr{k+k^2-4\vk\cdot\vP(1-\eta)}{[k-2\vk\cdot\vP(1-\eta)+k^2]^2}
 \label{resbound}\ea

\noi is similar to the bound obtained in \cite{Thom}. A compromise
choice

\ba
f_{\vk}=-[V_k+2\eta\vk\cdot\vP]/[k-2\vk\cdot\vP+k^2],\label{compchoice}
\ea

\noi eliminating all terms linear in  Bose operators $\aka$, $\ak$
in (\ref{finalHam}), is equally possible too, with the
corresponding self-consistency equation for $\eta$

\[ \eta P^2=
\dsum_{\vk}f^2_{\vk}\vk=\dsum_{\vk}\vk\cdot\vP[V_k+2\eta\vk\cdot\vP
]^2/[k-2\vk\cdot\vP+k^2]^2,\]

\noi which can be solved analytically. At the same time, the
simplest choice

 \[
f_k=-V_k/(k+k^2) \]

\noi seems to be the choice of preference, because the
technicalities of calculation of arbitrary order moments $\langle
0|{\mathcal H}^m(f)|0\rangle$ for this choice are exactly the same
as they were in the case $\vP=0$, i.e. based on the Wick theorem
exclusively and without involvement of any integrations over  wave
vectors more complicated and laborious than the integration
(\ref{exint}). Also, due to spherical symmetry of this choice,
several terms in the Hamiltonian (\ref{altfinalHam}) disappear,
thus making calculations at higher orders of the applied
variational method less laborious.

\section{Summary}

 It was shown that ground-state energy function $E_g(\alpha,\vP,k_0)$
 of the slow-moving acoustical polaron can be approximated
 from above by infinite convergent sequence of upper bounds applicable for arbitrary
 values of the electron-phonon interaction strength $\alpha$,  polaron total momentum
$\vP$ and limiting wave vector $k_0$. These bounds are provided by
the generalized variational method. Then, various experimentally
observable polaron characteristics of practical interest can be
derived from these bounds. The proposed algorithm for the
construction of the upper bounds is well suited for implementation
by means of modern programming and computational environments
destined for seamless fusion of analytical and numerical
computation within the same application, such as, for example,
{\em Mathematica} or {\em Maple}. The usage of the parallel
computing techniques is advisable and  would be highly
advantageous, too, due to the intrinsic nature of the algorithm
heavily relying on the Wick theorem and recursion relations for
massive analytic integrations over wave vectors. Actually, when
calculating each subsequent moment  $\langle 0|{\mathcal
H}^m(f)|0\rangle$, one has to calculate anew only the
contributions stemming from the connected graphs, because all
other contributions have already been calculated at the previous
stages of the calculations. For example,

\[
\langle 0|{\mathcal H}^3(f)|0\rangle=\langle 0|{\mathcal
H}(f)|0\rangle^3 +3  \langle 0|{\mathcal H}(f)|0\rangle\langle
0|{\mathcal H}^2(f)|0\rangle+   \langle\langle 0|{\mathcal
H}^3(f)|0\rangle \rangle, \nn\]

\noi where $\langle\langle ...\rangle\rangle$ stands for the
connected part. The proposed approach is in no way limited to the
acoustical polaron model considered above. It is rather universal
and, being so, applicable without any major alterations to a broad
range of polaron models of all sorts, including those ones
concerned with manifestations of various polaron-like phenomena in
quantum systems of lowered dimensions, such as quantum wells,
wires and dots, with or without external electric and/or magnetic
fields.

\section*{Acknowledgements}

N.N.B. would like to thank the Abdus Salam International  Centre
for Theoretical Physics, Trieste, Italy, for hospitality.

\begin{wrapfigure}{i}{0.5\textwidth}
\centerline{\includegraphics[width=0.46\textwidth]{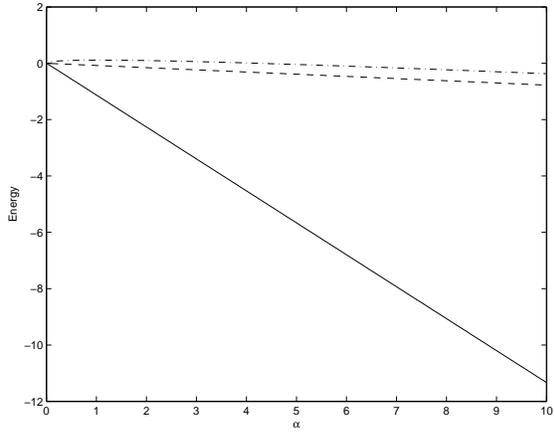}}
\caption{Upper bounds:  $E_{var}$, solid line; $E_W$, dashed line;
$E_{SC}$, dash-dotted line; $k_0=0.5$. } \label{fig1}
\end{wrapfigure}

\begin{wrapfigure}{i}{0.5\textwidth}
\centerline{\includegraphics[width=0.46\textwidth]{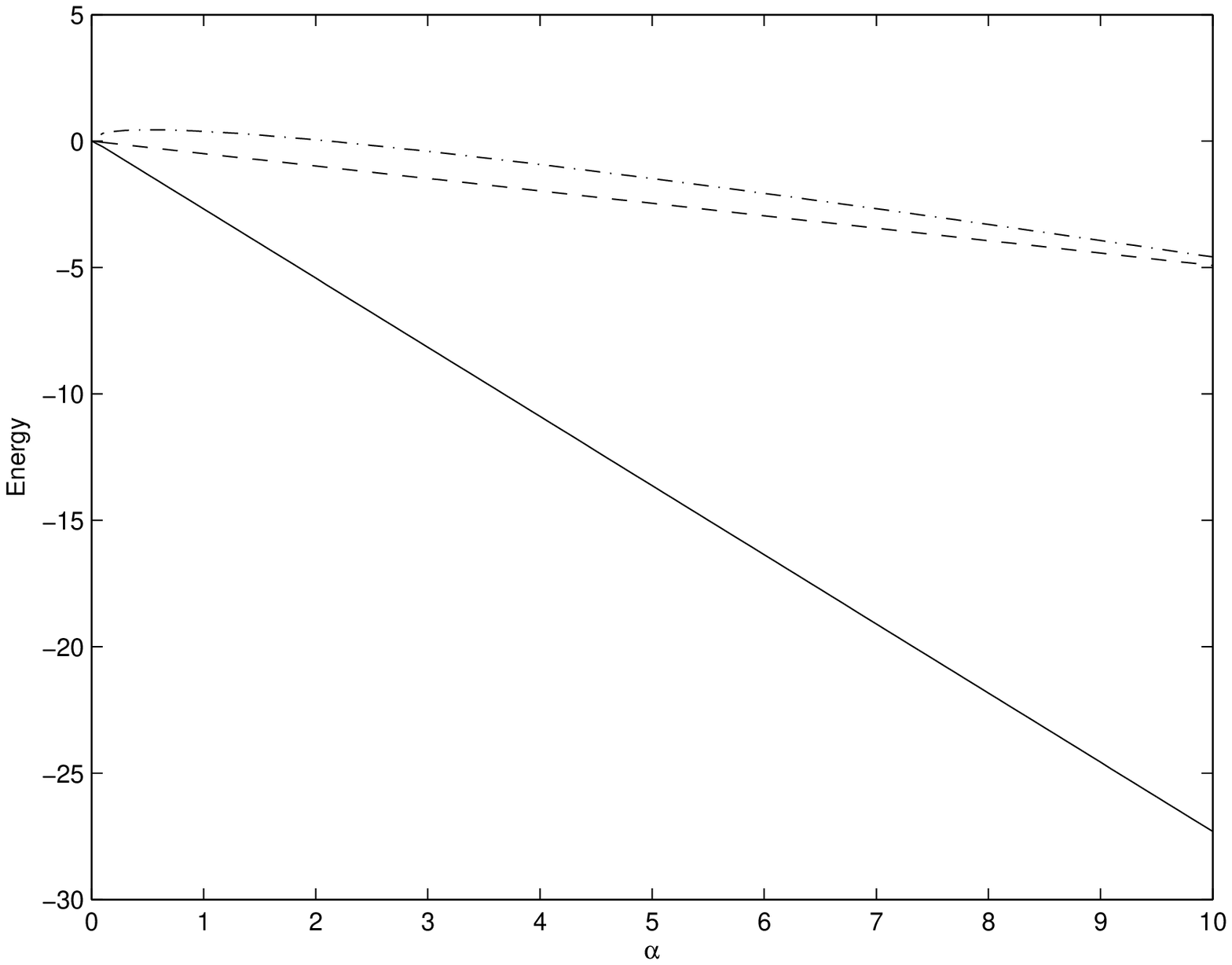}}
\caption{Upper bounds:  $E_{var}$, solid line; $E_W$, dashed line;
$E_{SC}$, dash-dotted line; $k_0=1$. } \label{fig2}
\end{wrapfigure}

\begin{wrapfigure}{i}{0.5\textwidth}
\centerline{\includegraphics[width=0.46\textwidth]{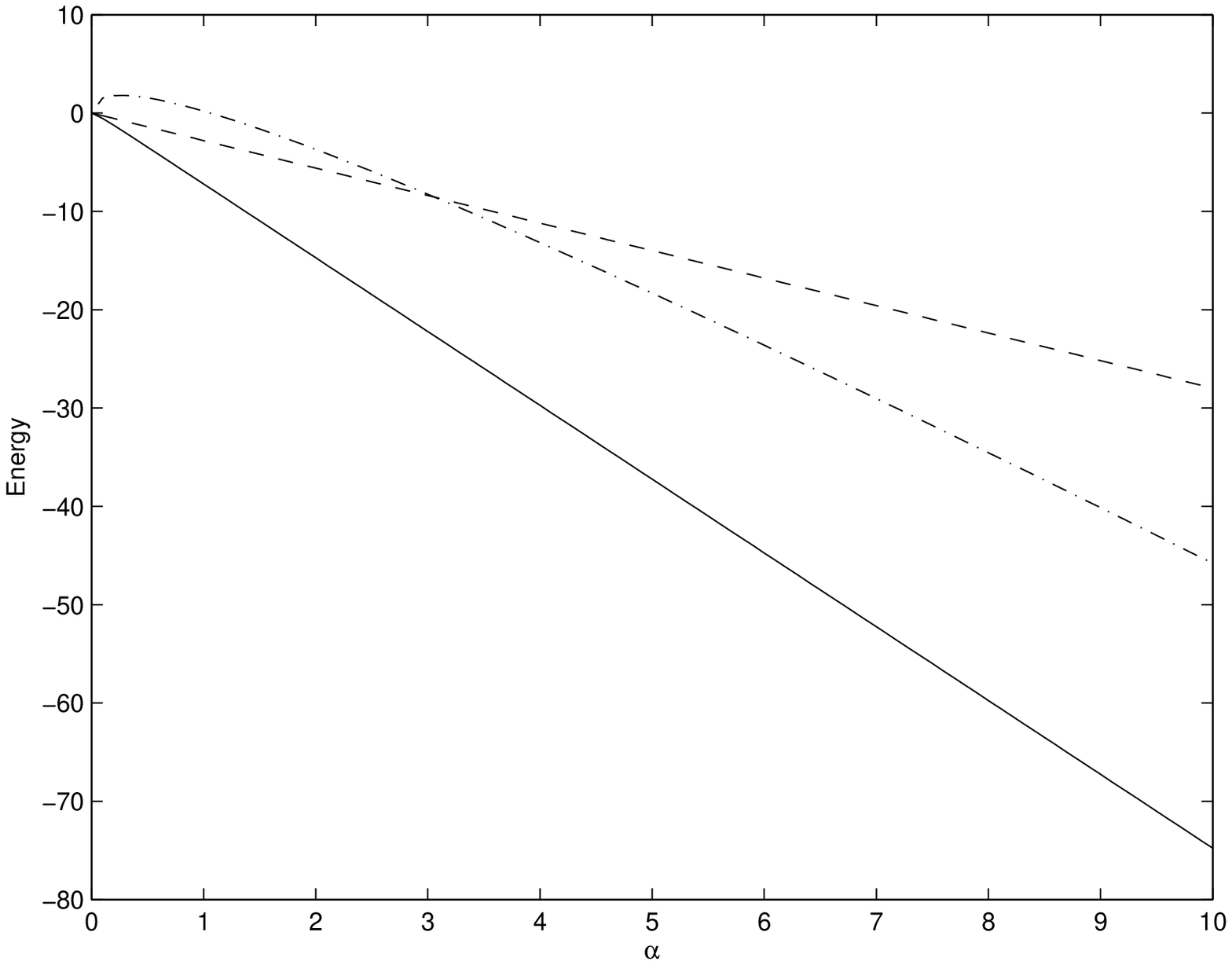}}
\caption{Upper bounds:  $E_{var}$, solid line; $E_W$, dashed line;
$E_{SC}$, dash-dotted line; $k_0=2$. } \label{fig3}
\end{wrapfigure}

\begin{wrapfigure}{i}{0.5\textwidth}
\centerline{\includegraphics[width=0.46\textwidth]{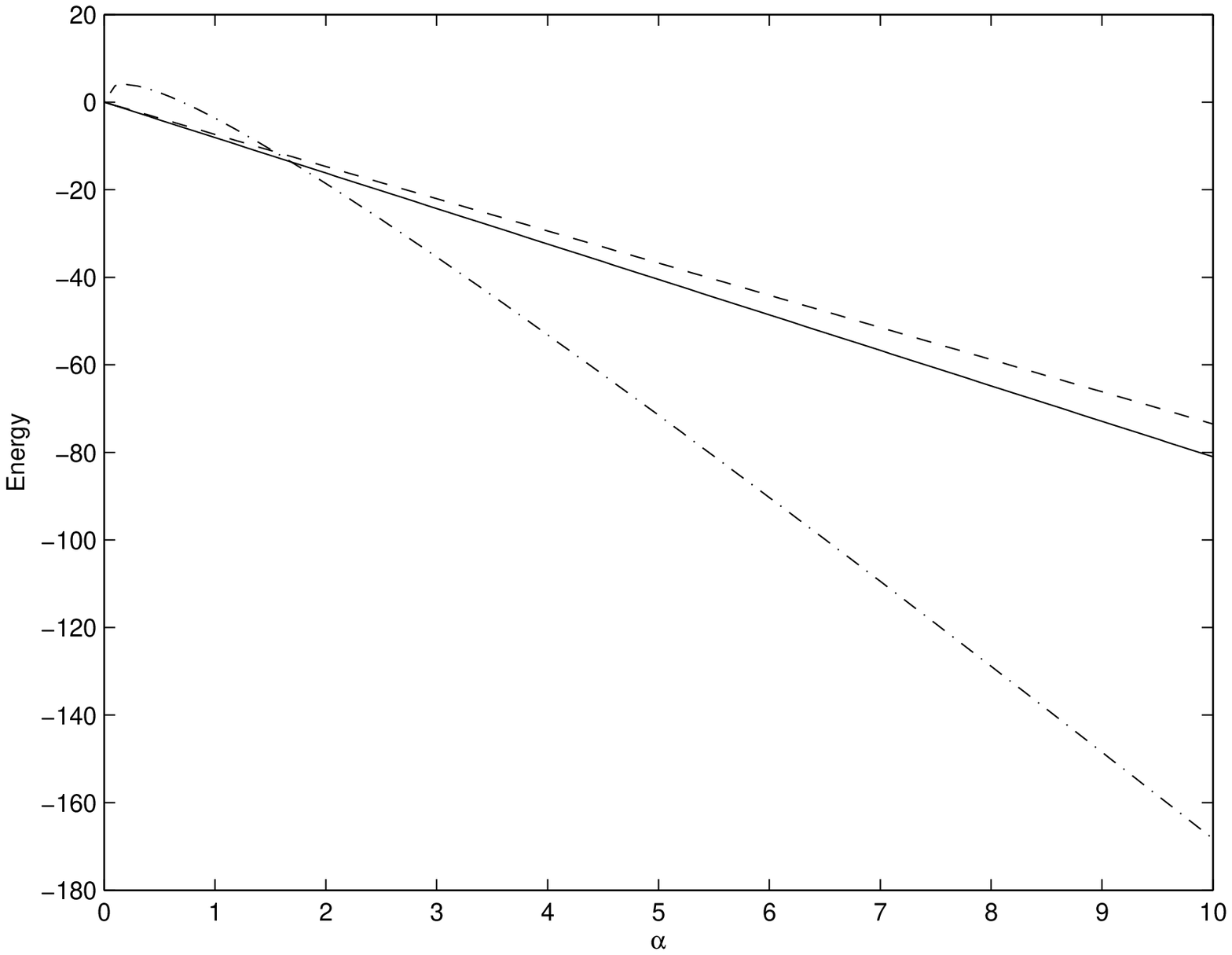}}
\caption{Upper bounds:  $E_{var}$, solid line; $E_W$, dashed line;
$E_{SC}$, dash-dotted line; $k_0=3$. } \label{fig4}
\end{wrapfigure}

 \label{last@page}
  \end{document}